\documentclass[aps,pre,twocolumn,amsmath,amsfonts,amssymb,floatfix,superscriptaddress,nofootinbib]{revtex4-1}
\usepackage{epsfig,amsmath,amssymb,graphics,color,calc}
\usepackage{mathtools}
\usepackage{xparse}

\newcommand{\ie}{\textit{i.e.}}

\begin{document}

\title{Critical behavior of the Anderson model on the Bethe lattice via a large-deviation approach}

\author{Giulio Biroli}
\affiliation{Laboratoire de Physique Statistique, Ecole Normale Sup\'erieure,
PSL Research University, 24 rue Lhomond, 75005 Paris, France}

\author{Alexander K. Hartmann}
\affiliation{Institut f\"ur Physik, Universit\"at Oldenburg, 26111 Oldenburg,
Germany}

\author{Marco Tarzia} 
\affiliation{LPTMC, CNRS-UMR 7600, Sorbonne Universit\'e, 4 Pl. Jussieu, F-75005 Paris, France}
\affiliation{Institut  Universitaire  de  France,  1  rue  Descartes,  75005  Paris,  France}

\begin{abstract}
We present a new large-deviation approach to investigate the critical properties of the Anderson model on the Bethe lattice close to the localization transition in the thermodynamic limit. Our method allows us to study accurately the distribution of the local density of states (LDoS) down to very small probability tails as small as $10^{-50}$ which are completely out of reach for standard numerical techniques. 
We perform a thorough analysis of the functional form and of the tails of the probability distribution of the LDoS which yields for the first time a direct, transparent, and precise estimation of the correlation volume close to the Anderson transition. Such correlation volume is found to diverge exponentially when the localization is approached from the delocalized regime, in a singular way that is in agreement with the analytic predictions of the supersymmetric treatment.
\end{abstract}

\maketitle

\section{Introduction}

After more than a half century, the subject of Anderson localization is still very much alive~\cite{fiftylocalization} as proved by the recent observations of Anderson localization of atomic gases in one dimension~\cite{aspect} and of classical sound elastic waves in three dimensions~\cite{localizationelastic}. On the theoretical side several questions remain open: Although there is by now a good understanding of the localization transition in low dimensional systems, culminating in a functional renormalization group analysis by a $2+\epsilon$ expansion~\cite{ludwig}, the behavior in high dimensions~\cite{largeD}, in particular the existence of an upper critical dimension and the relationship with Bethe lattice analysis~\cite{abou}, is still an issue. Recently, there has been a renewal of interest on this problem because of its relationship with Many-Body localization (MBL)~\cite{BAA}. This is a fascinating new kind of phase transition between a low temperature non-ergodic phase---a purely quantum glass---and a high temperature ergodic phase~\cite{Gornyi2005,Altman2015Review,Nandkishore2015,AbaninPapic2017,AletLaflorencie2018,Abanin2019RMP}. This phenomenon has been argued to take place for several disordered isolated interacting quantum systems, 
and can be thought of as localization in the Fock space of Slater determinants, which play the role of lattice sites in a disordered Anderson tight-binding model. A paradigmatic representation of this transition~\cite{A97,BAA,jacquod,wolynes,scardicchioMB,roylogan,mirlinreview} is indeed (single-particle) Anderson localization on a very high dimensional hierarchical lattice, which for spinless electrons consists in an $N$-dimensional hyper-cube (where $N\gg1$ is the number of sites of the lattice system). 
Although the analogy between MBL and Anderson localization on the Bethe lattice involves several drastic simplifications (e.g. the correlation between random energies are neglected as well as the specific structure of the Hilbert space),  it is very useful to obtain a qualitative understanding of the problem~\cite{mirlinreview,dinamica,tikhonovmirlinMBL,Biroli2020}. 

Localization had an impact on several fields, in particular Random Matrices and Quantum Chaos. As a matter of fact, in the delocalized phase the level statistics is described by random matrix theory and generally corresponds to the Gaussian Orthogonal Ensemble (GOE), whereas instead in the localized phase is determined by Poisson statistics because wave-functions close in energy are exponentially localized on very distant sites and hence do not overlap; thus, contrary to the GOE case, there is no level-repulsion and eigenenergies are distributed similarly to random points thrown on a line. 

The relationship with quantum chaos goes back to the Bohigas-Giannoni-Schmidt conjecture, which states that the level statistics of chaotic (or ergodic) systems is given by random matrix theory, whereas integrable systems instead are characterized by Poisson statistics~\cite{BGS}. This result can be fully worked out and understood in the semi-classical limit~\cite{berry,altshulerchaos}: for a quantum chaotic system, in the $\hbar \rightarrow 0$ limit, wave-functions at a given energy become uniformly spread over the micro-canonical hyper-surface of the configuration space. They are fully delocalized as expected for an ergodic classical system that covers regions with same energy uniformly. Instead, quantum non-ergodic models, such as integrable systems, are characterized by Poisson statistics and localized wave-functions. All those results support a general relationship between delocalization--GOE statistics--ergodicity (similarly between localization--Poisson statistics--lack of ergodicity).

However, in the last decade several numerical studies~\cite{noi,scardicchio1,ioffe1,ioffe3,bera2018,detomasi2020,refael2019} have been performed for the Anderson model on the \emph{Bethe lattice}, in fact, on \emph{Random-Regular Graphs} (RRG), with $N$ nodes and a
parameter $W$ controlling the strength of the local disorder. This is
a class of random lattices that have locally a tree-like structure but do not have boundaries, see below for a precise definition. The results have suggested the possibility of the existence of an intermediate delocalized but non-ergodic phase characterized by multifractal eigenfunctions in a broad disorder range preceding the localization transition,  as first suggested in~\cite{A97}. The arguments in favour of this scenario rely mostly on numerical extrapolations of results obtained from Exact Diagonalization (ED) of large but finite samples, and the existence of such a phase in the thermodynamic limit has been strongly questioned during recent years~\cite{mirlin,lemarie,tikhonov2019,biroli2018,levy,mirlinreview,metz}.

Although the possibility of such multifractal delocalized phase is clearly very intriguing, especially due to its relationship with MBL~\cite{A97},
it appears to be in explicit conflict with the analytical predictions based on the supersymmetric approach for the Anderson model on sparse random graphs~\cite{SUSY,fyod,mirlin1994,Zirn}.
Moreover, recent numerical investigations based on the finite-size scaling of the spectral and the 
wave-functions statistics on the delocalized side of the Anderson model on RRG~\cite{mirlin,tikhonov2019,biroli2018} and similar sparse random lattices~\cite{levy,lemarie} provided strong indications against the existence of a truly intermediate non-ergodic extended phase. These investigations have highlighted a non-monotonous behavior of the observables as a function of the system size on the delocalized side of the transition, which can be explained in terms of (i) the presence of a characteristic scale which diverges exponentially fast approaching the transition and is already very large far from it~\cite{lemarie,levy,biroli2018,mirlin,tikhonov2019}; (ii) the localized nature of the critical point in the limit of infinite dimension~\cite{largeD,fyod,efetov,Zirn}. The combination of these two elements produce {\it dramatic and highly non-trivial finite size effects} even very far from the critical point, and give rise to a strong non-ergodic behavior in a crossover region where the \emph{correlation volume} $N_c(W)$ is larger than the accessible system sizes. (On the contrary, there is by now a general consensus on the fact that the delocalized phase of the Anderson model on the loop-less Cayley is genuinely multifractal~\cite{garel,cayley,Biroli2020}).

Note that the thorough characterization of such crossover regime has not only an academic interest but has also some important practical implications. In fact the crossover scale turns out to be so large even far below the localizaiton transition that the multifractal exponents associated to the spectral statistics appear to be independent on the system size $N$ in a broad range of sizes smaller than $N_c$, producing an effective non-ergodic behavior on several decades of length and timescales~\cite{dinamica,bera2018,detomasi2020,Biroli2020}. Yet, a precise characterization of the correlation volume, in particular from the numerical point of view,  remains elusive. Direct numerical simulations would need to focus on intractably large system sizes. The 
 Anderson transition on tree-like lattices offers however an alternative route, since it allows for an exact solution~\cite{abou,tikhonov2019,biroli2018,SUSY,fyod,mirlin1994,Zirn,ourselves,aizenmann,semerjian,tikhcrit,parisi,metz}. This can be obtained in terms of the exact self-consistent equations for the Green's functions (in the thermodynamic limit), which allow to establish the transition point and the corresponding critical behavior. However, even this approach suffers from the dramatic increase of the correlation volume, which controls the cutoff of the probability distribution of the imaginary part of the Green's function (\ie, the local density of states (LDoS))~\cite{SUSY,fyod,mirlin1994,tikhonov2019,tikhcrit}. Since $N_c(W)$ is so large even far away from the transition, the cutoff occurs in the far-tails of the distribution which cannot be properly sampled with standard numerical techniques such as the population dynamics algorithm even using huge populations~\cite{tikhcrit}. Here, we solve this problem by putting forward a novel large-deviation technique which allows one to sample very accurately the tails of the probability distribution of the LDoS down to extremely small probabilities, and highlight with great accuracy the crossover scale and its critical behavior.


The main conclusions of our analysis fully confirm the predictions of the supersymmetric approach~\cite{tikhonov2019,SUSY,fyod,mirlin1994,Zirn} and are compatible with a correlation volume which diverges exponentially fast as the Anderson localization is approached, as $N_c(W) \approx A \, e^{c/(W_L-W)^\nu}$, with $\nu=1/2$ and $W_L$ being the critical disorder strength.

The paper is organized as follows. In the next section we introduce the model and briefly review previous results and studies. In Sec.~\ref{sec:numerics} we present some recent numerical results of the spectral statistics obtained from ED of the Anderson model on the RRG of finite size. In Sec.~\ref{sec:BP} we describe the new large deviation approach to sample efficiently the tails of the distributions of the Green's functions and determine accurately the correlation volume close to $W_L$. Finally, in Sec.~\ref{sec:conclusions} we discuss the physical implications of our results, providing some concluding remarks and perspectives for future work.

\section{Model and State of the Art} \label{sec:model}

The model we focus on consists in non-interacting spinless electrons in a disordered potential: 
\begin{equation} \label{eq:H}
{\cal H} = - t \sum_{\langle i,j \rangle} \left( c_i^{\dagger} c_j
+ c_j^{\dagger} c_i \right ) - \sum_{i=1}^N \epsilon_i \, c_i^\dagger c_i \, ,
\end{equation}
where the first sum runs over all the nearest neighbors sites of the lattice, the second sum runs over all $N$ sites; $c_i^\dagger$, $c_i$ are fermionic creation and annihilation operators, and $t$ is the hopping kinetic energy scale, which we take equal to $1$. The on-site energies $\epsilon_i$ are i.i.d. random variables uniformly distributed in the interval $[-W/2,W/2]$:
\begin{equation} \label{eq:peps}
p(\epsilon) = U \left(-\frac{W}{2},\frac{W}{2} \right) \equiv \frac{1}{W} \, \theta \! \left ( \frac{W}{2} - | \epsilon | \right) \, .
\end{equation}

As anticipated in the introduction, the lattice that we consider is a  $(k+1)$-RRG, {\it i.e.}, a lattice chosen uniformly at random among all graphs of $N$ sites where each of the sites has connectivity $k+1$.  The properties of such random graphs have been extensively studied (see Ref.~\cite{wormald} for a review). A RRG can be essentially viewed as a finite portion of a tree wrapped onto itself. It is known in particular that for large number of sites any finite portion of such a graph is a tree with a probability going to one as $N \to \infty$, and that the RRG has large loops of typical length of order $\ln N$~\cite{wormald}. 

The model~(\ref{eq:H}) is then a sum of two random matrices, ${\cal H} = {\cal C} + {\cal D}$: ${\cal C}$ is the connectivity matrix of the RRG, ${\cal C}_{ij} = -t$ if sites $i$ and $j$ are connected and zero otherwise. ${\cal D}$ is the diagonal matrix corresponding to the on-site random energies,  ${\cal D}_{ij} = \epsilon_i \delta_{ij}$. It is known from previous studies that the former ensemble of sparse random matrices belongs to the GOE universality class (with fully delocalized eigenvectors)~\cite{RRG-GOE,Bauerschmidt}, while the latter is described by definition by Poisson statistics (with fully localized eigenvectors).

Localization on the RRG was first studied by Abou-Chacra, Anderson and Thouless~\cite{abou} and then later by many others, see~\cite{noi,scardicchio1,ioffe1,ioffe3,bera2018,detomasi2020,refael2019,fyod,Zirn,Verb,ourselves,aizenmann,semerjian,tikhonov2019,tikhcrit,parisi,metz,biroli2018,mirlin1994,SUSY,gabriel} and Refs.~therein. Many similarities, but also few important differences, with the $3d$ behavior have been found. The differences mainly concern the critical properties. Contrary to the finite-dimensional case, the critical behavior is not power-law-like but instead exponential, \ie, one finds essential singularities approaching the localization transition from the delocalized regime~\cite{fyod,efetov,Zirn,SUSY,mirlin1994,tikhonov2019}. Moreover, the inverse participation ratio (IPR), defined as 
$\langle \sum_{i=1}^N |\psi_\alpha (i) |^4 \rangle$, is found to have a discontinuous jump at the transition from a $O(1)$ toward a $1/N$ scaling~\cite{SUSY}, instead of being continuous at the transition. Arguments based on supersymmetric field theory indicate that the level statistics should display a transition from GOE to Poisson statistics concomitant with the localization transition~\cite{fyod,SUSY} (see also Ref.~\cite{metz}). However, the first numerical studies didn't fully support this claim~\cite{noi,berkovits}. Moreover, the arguments of~\cite{A97} indicates that the two transitions might actually not coincide. As discussed above, the possibility of the existence of an intermediate phase, which is delocalized and yet still not ergodic, were first suggested in~\cite{noi}. These findings triggered a lot of activity. In Ref.~\cite{scardicchio1}, based on the numerical extrapolation of the spectrum of fractal dimensions of finite size systems, it was conjectured that the eigenstates are multifractal in the whole delocalized phase. More recently, the authors of Refs.~\cite{ioffe1,ioffe3} combined EDs and semi-analytical calculations to claim the existence of the intermediate non-ergodic but delocalized phase in a broad disorder strength $W_E < W < W_L$. These claims have been questioned by the numerical investigations of Refs.~\cite{mirlin,levy,lemarie,tikhonov2019,biroli2018,metz} which analyzed the level and eigenfunction statistics on the delocalized side of the Anderson transition on the RRG and similar sparse random lattices, and unveiled the existence of very strong finite size effects with a characteristic crossover scale $N_c(W)$ associated to a pronounced non-monotonous behavior of the observables as a function of $N$, and which diverges exponentially fast as the localization transition is approached.  The origin of the non-monotonicity has been traced back to the localized nature of the Anderson critical point in the limit of infinite dimensions~\cite{largeD,fyod,efetov,Zirn}: For $N \ll N_c$ the system flows towards the Anderson transition fixed point, whose properties on the RRG are analogous to the localized phase, whereas for $N \gg N_c$ the system approaches the $N \to \infty$ ergodic behavior. The conclusion of these investigations 
are thus that the system is ergodic in the whole delocalized phase, but is characterized by dramatic and non-trivial finite-size effects even very far from the critical point, giving rise to an apparent non-ergodic behavoir in a crossover region where the correlation volume is larger than the accessible system sizes. Nonetheless, as explained in the introduction, a precise characterization of the correlation volume $N_c$ is still missing. 

In the following, without loss of generality, we focus on the $k=2$ case ({\it i.e.}, total connectivity $k+1 = 3$) and on the middle of the spectrum, $E=0$. Previous studies of the transmission properties and dissipation propagation determined that the localization transition takes place at $W_L \approx 18.2$~\cite{abou,garel,ourselves,tikhcrit}, while previous analysis of the spectral properties have suggested the presence of the non-ergodic delocalized phase in the range $10 \approx W_E < W < W_L$~\cite{noi,ioffe1,ioffe3}.

\section{Exact diagonalization on the RRG} \label{sec:numerics}

The purpose of this section is to show results, in agreement with the recent literature~\cite{mirlin,levy,lemarie,tikhonov2019,biroli2018}, that support the presence of the correlation volume $N_c(W)$ and its very fast increase. In particular we shall focus on numerical results for the level statistics of the Anderson model on the RRG which 
unveil the non-monotonic behavior of the relevant observables. 
These results are obtained from EDs of the Hamiltonian~(\ref{eq:H}) on the RRG for several system sizes $N=2^n$, from $n=6$ to $n=15$, and for several values of the disorder strength $W$ on the delocalized side of the Anderson transition in the disorder range where previous studies have suggested the possibility of the existence of a multifractal delocalized phase~\cite{ioffe1,ioffe3}, $W_E < W < W_L$. 
For each value of $N$ and $W$, we average over both the on-site quenched disorder and on RRG realizations, taking (at least) $2^{22-n}$ different samples. Since we are interested in $E=0$, we only focused on $1/8$ of the eigenstates centered around the middle of the band (we have checked that taking $1/16$ or $1/32$ of the states does not alter the results, but yields a poorer statistics).

We study the statistics of level spacings of neighboring eigenvalues: $s_\alpha = E_{\alpha+1} - E_\alpha \ge 0$, where $E_\alpha$ is the energy of the $\alpha$-th eigenstate in the sample. In the delocalized regime level crossings are forbidden. Hence the eigenvalues are strongly correlated and the level statistics is expected to be described by Random Matrix Theory (more precisely, several results support a general relationship between delocalization and the Wigner's surmise of the GOE). Conversely, in the localized phase wave-functions close in energy are exponentially localized on very distant sites and do not overlap. Thus there is no level-repulsion and eigenvalues should be distributed similarly to random points thrown on a line (Poisson statistics). In order to avoid difficulties related to the unfolding of the spectrum, we follow~\cite{huse} and measure the ratio of adjacent gaps, 
\[
r_\alpha = \frac{\min \{ s_\alpha, s_{\alpha+1} \}}{\max \{ s_\alpha, s_{\alpha+1} \}} \, ,
\]
and obtain the probability distribution which displays a universal form depending on the level statistics~\cite{huse}. In particular $\langle r \rangle$ is expected to converge to its GOE and Poisson counterpart in the extended and localized regime~\cite{Pr-GOE}, allowing to discriminate between the two phases as $\langle r \rangle$ changes from $\langle r \rangle_{\rm GOE} \simeq 0.53$ to $\langle r \rangle_P \simeq 0.39$ respectively.

The GOE-Poisson transition can also be captured by correlations between nearby eigenstates such as the mutual overlap between two subsequent eigenvectors, defined as:
\[
q_m = \sum_{i=1}^N | \psi_\alpha (i)  | | \psi_{\alpha+1} (i) | \, .
\]
In the GOE regime the wave-functions amplitudes are i.i.d. Gaussian random variables of zero mean and variance $1/N$~\cite{porter-thomas}, hence $\langle q \rangle$ converges to $\langle q \rangle_{\rm GOE} = 2/\pi$. Conversely in the localized phase two successive eigenvector are typically peaked around very distant sites and do not overlap, and therefore $\langle q \rangle_{P} \to 0$ for $N \to \infty$. At first sight this quantity seems to be related to the statistics of wave-functions' coefficients rather than to energy gaps. Nonetheless, in all the random matrix models that have been considered in the literature so far, one empirically finds that $\langle q \rangle$ is directly associated to the statistics of gaps between neighboring energy levels~\cite{notaRP}. 

\begin{figure}
 \includegraphics[angle=0,width=0.51\textwidth]{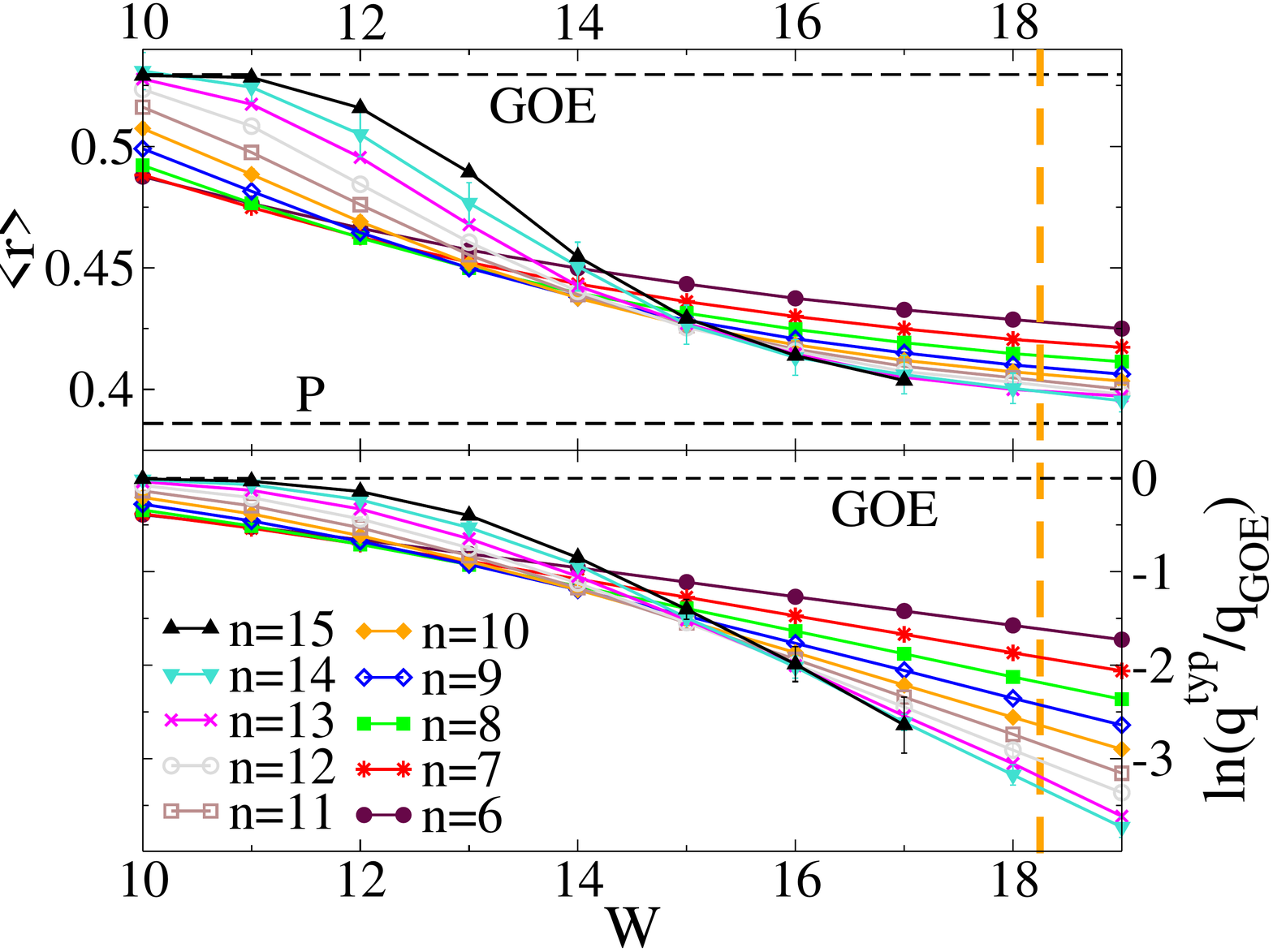}
 \caption{\label{r-q} (color online)
 $\langle r \rangle$ (upper panel) and $\ln (q^{\rm typ}/q_{\rm GOE})$ (lower panel) as a function of the disorder $W$ for several system sizes $N=2^n$ with $n$ from $6$ to $15$. The horizontal dashed lines correspond to the reference GOE and Poisson asymptotic values. The vertical orange dashed line spots the position of the Anderson localization transition, $W_L \approx 18.2$~\cite{tikhcrit}. }
 \end{figure}
 
\begin{figure}
 \includegraphics[angle=0,width=0.5\textwidth]{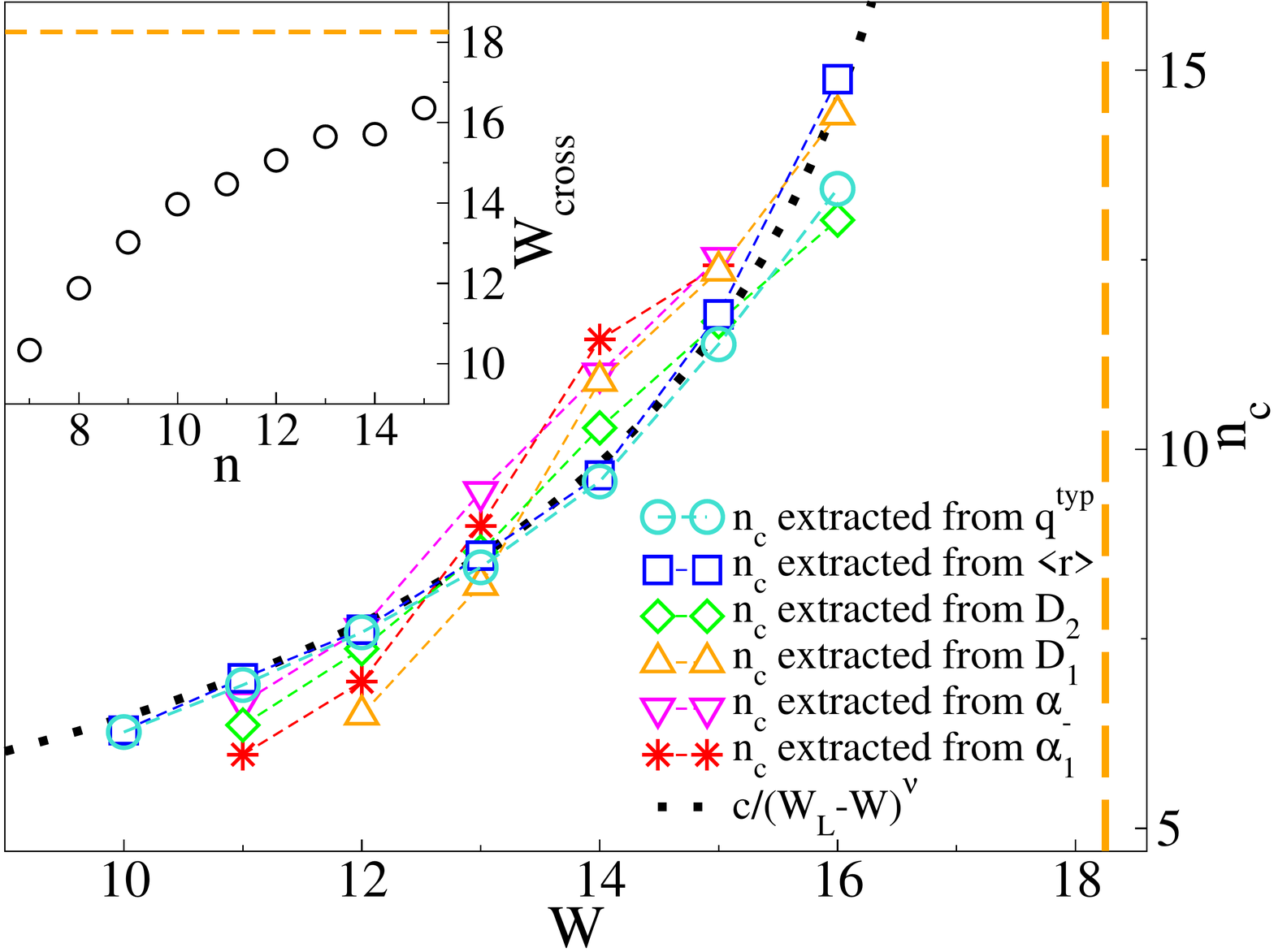}
 \caption{\label{nc} (color online)
 Main panel: Characteristic crossover scales $n_c (W)=\log_2 N_c(W)$ extracted from different
	observables related to the level statistics ($\langle r \rangle$, squares, and $q^{\rm typ}$, circles) and to the statistics of the wavefunctions' amplitude (the fractal dimension $D_2$, diamonds, the fractal dimension $D_1$,  up triangles, the edge $\alpha_-$ of the support of the multifractal spectrum $f(\alpha)$,  down triangles, and the point $\alpha_1$ where $f(\alpha_1) = \alpha_1$ and $f^\prime (\alpha_1)=1$, stars). See Ref.~\cite{biroli2018} for more details. The black dotted curve is a fit of the form $n_c (W) \propto c/(W_L-W)^\nu$ with $c \approx 20$ and $\nu \approx 0.6$. Inset: Evolution with $n$ of the crossing point of the curves of $q^{\rm typ}(W)$ of Fig.~\ref{r-q} for two subsequent system sizes.}
 \end{figure}

In Fig.~\ref{r-q} we show the behavior of the average value of the ratio of adjacent gaps, $\langle r \rangle$, and of (the logarithm of) the typical value of the mutual overlap between subsequent eigenvectors, $q^{\rm typ} = e^{\langle \ln q \rangle}$, as a function of the disorder $W$, for several system sizes $N=2^n$, with $n$ from $6$ to $15$. As expected, for small (resp. large) enough disorder we recover the universal values $\langle r \rangle_{\rm GOE} \simeq 0.53$ and $q_{\rm GOE}^{\rm typ} = 2 / \pi$ (resp. $\langle r \rangle_{P} \simeq 0.39$ and $q_P^{\rm typ} \to 0$) corresponding to GOE (resp. Poisson) statistics. However, as pointed out in~\cite{noi} the different curves corresponding to different values of $N$ cross much before the localization transition, occurring at $W_L \approx 18.2$, as indicated by the vertical dashed line in the plot. This behavior was interpreted in terms of an intermediate delocalized but non-ergodic phase~\cite{noi}. Nevertheless, analyzing carefully  the data, we realized that the crossing point is in fact slowly but systematically  drifting towards larger values of $W$ as $N$ is increased (see inset of Fig.~\ref{nc}), as also observed~\cite{mirlin,levy,biroli2018}. 
  
\begin{figure}
 \includegraphics[angle=0,width=0.5\textwidth]{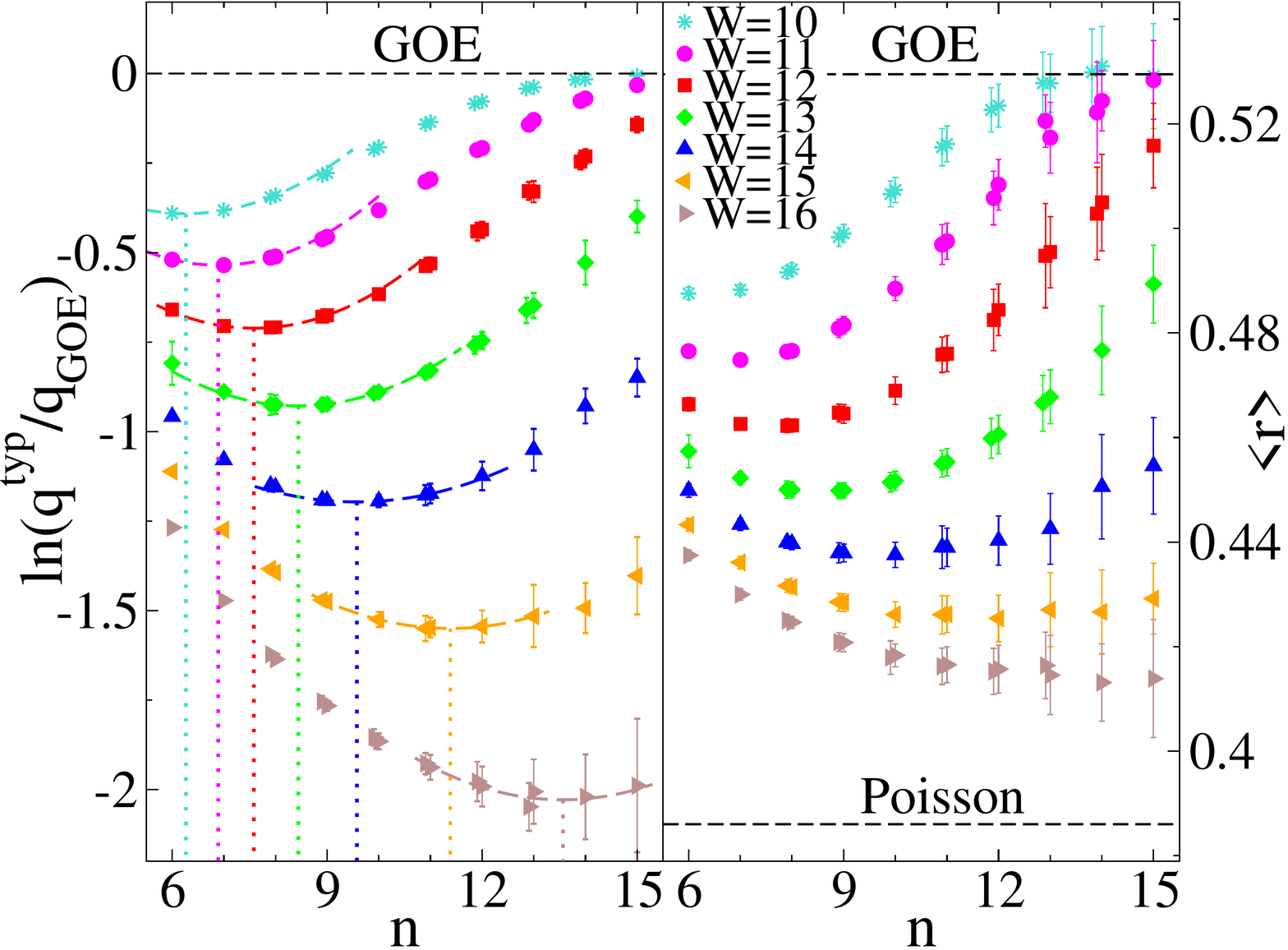}\caption{\label{minimo2}
 (color online) $\ln (q^{\rm typ}/q_{\rm GOE})$ (left panel) and $\langle r \rangle$ (right panel) as a function of $n=\log_2 N$ for $W=10,11,\ldots,16$. The data show the non-monotonic  behavior of $q^{\rm typ}$ and $\langle r \rangle$. The position of the minimum $n_c(W)$ extracted from $q^{\rm typ}(W)$, corresponding to the volume $N_c(W)=2^{n_c(W)}$, is represented by the vertical dotted lines.}
 \end{figure}
 
This is clearly unveiled by Fig.~\ref{minimo2}, where we plot the behavior of $q^{\rm typ}$ and $\langle r \rangle$ as a function of $n = \log_2 N$, for several values of the disorder belonging to the range where the curves of $\langle r \rangle$ and $q^{\rm typ}$ for different $n$ cross, \ie, $10 \lesssim W \lesssim 16$. One indeed observes that in this region $q^{\rm typ}$ and $\langle r \rangle$ become non-monotonic functions of $n$. The position of the minimum of $q^{\rm typ}$ (highlighted by dashed vertical lines in the left panel of Fig.~\ref{minimo2}) naturally defines a characteristic system size, $N_c(W) = 2^{n_c (W)}$, governing the crossover from Poisson to GOE statistics (on the scale of the mean level spacing): For $N<N_c(W)$ one has indeed that $q^{\rm typ}$ decreases as the system size is increased, as expected for localized wave-functions, whereas for $N>N_c(W)$ it is an increasing function of $n$ and eventually converges to the GOE universal value. The same non-monotonic behavior as a function of the system size is found for $\langle r \rangle$ (right panel of Fig.~\ref{minimo2}), as well as for many other observables related to the wave-functions' statistics, such as the IPR and the multifractal spectrum, as previously observed in Refs.~\cite{mirlin,levy,biroli2018}. 

These results indicate the emergence of a unique characteristic scale which controls the transition from a phase characterized by Poisson statistics, localization and lack-of-ergodicity to one displaying GOE statistics, delocalization and ergodicity for the Anderson model on RRGs of finite size. 
This is confirmed by the main panel of Fig.~\ref{nc}, where we plot the characteristic crossover scales, $n_c (W)$, extracted from the different probes related to both the statistics of the gap and the statistics of the wavefunctions' amplitudes, showing that, within our numerical accuracy, they all yield a very similar dependence on the disorder strength $W$ (see Ref.~\cite{biroli2018} for more details).

As anticipated above, the non-monotonic behavior has been interpreted in~\cite{mirlin} in terms of the nature of the Anderson critical point on the RRG, which has properties similar to that of the localized phase~\cite{largeD,fyod,efetov,Zirn}, with critical level statistics of Poisson form and strongly localized critical wave-functions. The observables of systems of size $N \ll N_c(W)$ would then first flow upon increasing $N$ towards the critical values, which tend, for $d \to \infty$, to the ones of the localized phase~\cite{largeD,fyod,efetov,Zirn}. Then, when $N$ becomes larger than the correlation volume $N_c$, the observables flow towards their standard values in the delocalized, fully ergodic, phase.

The black dotted curve of Fig.~\ref{nc} shows a fit of the data of the form $n_c \propto c/(W_L - W)^\nu$, 
implying an exponential divergence of the correlation volume at the transition point. However, our numerical data are clearly too far from $W_L$ to obtain an accurate estimation of $\nu$. Yet, the value of the exponent is not too far from the one predicted by the supersymmetric analysis, $\nu = 1/2$~\cite{SUSY,fyod,mirlin1994,Zirn}. 
In the next section we put forward a new large-deviation approach which allows one to access the crossover scale from the solution of the self-consistent equations for the Green's functions in the thermodynamic limit, providing a much more stringent test of the analytic predictions.



\section{Self-consistent iteration equations for the Green's functions and large deviation method} 
\label{sec:BP}

As discussed in the introduction, the Anderson model on tree-like structures allows for an exact solution in the limit of infinite lattices~\cite{abou,tikhonov2019,SUSY,fyod,mirlin1994,Zirn,ourselves,aizenmann,semerjian,tikhcrit,metz}, which yield the probability distribution function of the diagonal elements of the resolvent matrix, defined as ${\cal G} (z) =  ({\cal H} - z {\cal I}  )^{-1}$.

In order to obtain the recursive equations, the key objects are the so-called {\it cavity} Green's functions, $G_{i \to j} (z) = [({\cal H}_{i \leftrightarrow j} - z {\cal I}  )^{-1}]_{ii}$, \ie, the diagonal elements on site $i$ of the resolvent matrix of the modified Hamiltonian ${\cal H}_{i \leftrightarrow j}$ where the edge between the site $i$ and one of its neighbors $j$ has been removed. 

Take a given site $i$ and its neighbors $\{l_1, \ldots, l_{k+1} \}$ living on an infinite tree. If one removes the site $i$ from the graph, then the sites $\{ l_1, \ldots, l_{k+1} \}$ are uncorrelated, since the lattice would break in $k+1$ semi-infinite disconnected branches. One then obtains (e.g., by direct Gaussian integration or using the block matrix inversion formula) the following iteration relations for the cavity Green's functions~\cite{abou}:
\begin{equation} \label{eq:recursion}
G_{i \to l_m}^{-1} (z) = - \epsilon_i - z -
t^2 \!\!\!\! \sum_{l_j \in \partial i / l_m}  \!\!\!\! G_{l_j \to i} (z) \, ,
\end{equation}
where $l_m$ with $m=1,\ldots,k+1$ denote the excluded neighbor of $i$, 
$z=E + {\rm i} \eta$, $\eta$ is an infinitesimal imaginary regulator which smoothens out the pole-like singularities in the right hand sides, $\epsilon_i$ is the on-site random energy taken from the  distribution~(\ref{eq:peps}), and $\partial i/l$ denotes the set of all $k+1$ neighbors of $i$ except $l$. (Note that for each site with $k+1$ neighbors one can define $k+1$ cavity Green's functions and $k+1$ recursion relations of this kind.) After that the solution of Eqs.~(\ref{eq:recursion}) has been found, one can finally obtain the diagonal elements of the resolvent matrix of the original problem on a given site $i$ as a function of the cavity Green's functions for all the neighboring sites~\cite{ourselves}:
\begin{equation} \label{eq:recursion_final}
{\cal G}_i^{-1} (z) = - \epsilon_i - z - t^2 \!
\sum_{l_j \in \partial i }  \! G_{l_j \to i} (z) \, .
\end{equation}
In the following we will focus on the middle of the spectrum ($E=0$) and set $t=1$.

The statistics of the diagonal elements of the resolvent gives---in the $\eta \to 0^+$ limit---the spectral properties of $\mathcal{H}$. In particular, the probability distribution of the LDoS at energy $E$ is given by:
\begin{equation} \label{eq:LDoS}
	\begin{aligned}
		\rho_i & = \sum_\alpha | \psi_\alpha(i)  |^2 \, \delta ( E - E_\alpha ) 
		=\lim_{\eta \to 0^+}  \frac{1}{\pi} \, {\rm Im} {\cal G}_i (z) \, ,
	\end{aligned}
\end{equation}
from which the average Density of States (DoS) is simply given by $\rho = (1/N) \sum_i \rho_i = 1/(N \pi) {\rm Tr} \, {\rm Im} {\cal G}$. 

Note that, however, on finite RRGs when site $i$ is removed from the graph, the neighbors $\{l_1, \ldots, l_{k+1} \}$ are not truly decoupled, since they are still connected by some (typically large) loop present somewhere in the system.  Since the average size of the loops scales as $\ln N$~\cite{wormald}, it is reasonable to expect that Eqs.~(\ref{eq:recursion}) and~(\ref{eq:recursion_final}) become asymptotically exact in the thermodynamic limit as the cavity Green's functions on sites  $\{l_1, \ldots, l_{k+1} \}$ become uncorrelated in absence of site $i$ if the typical length of the loops which connect them is larger than the correlation length. This has been in fact proven rigorously in Ref.~\cite{bored} using the local convergence of RRGs to  Cayley trees.

Since the Green's functions $G_{i \to j}$ and ${\cal G}_{i}$ are random variables, Eqs.~(\ref{eq:recursion}) and (\ref{eq:recursion_final}) naturally lead to functional equations on their probability distribution $Q (G)$ and $P({\cal G})$. From Eq.~(\ref{eq:recursion}) one first gets the self-consistent functional equation for the probability distributions of the cavity Green's functions in the $N \to \infty$ limit (averaged over the on-site disorder and on different realizations of the random lattice):
\begin{equation} \label{eq:PGcav}
\begin{split}
Q (G) &= \! \int \! \textrm{d}p(\epsilon) \prod_{l=1}^k \textrm{d} Q (G_l) \, 
\delta \! \left ( \! G^{-1} \! + \epsilon + z  + \sum_{i=1}^k G_l \! \right) \, ,
\end{split}
\end{equation}
where $p(\epsilon)$ is the probability distribution of the on-site random energy, Eq.~(\ref{eq:peps}). Once the fixed point of Eq.~(\ref{eq:PGcav}) is obtained, using Eq.~(\ref{eq:recursion_final}) one can compute the probability distribution of the diagonal elements of the resolvent:
\begin{equation} \label{eq:PG}
\begin{split}
	P ({\cal G}) &= \! \int \! \textrm{d}p(\epsilon) \prod_{l=1}^{k+1} \textrm{d} Q (G_l) \, 	\delta \! \left ( \! {\cal G}^{-1} \! + \epsilon + z + \sum_{l=1}^{k+1} G_l \! \right) \, .
\end{split}
\end{equation}
This set of functional equations can be solved numerically with an arbitrary degree of precision using a population dynamics algorithm~\cite{abou,ourselves,ioffe1,ioffe3,PopDyn,tikhcrit,lemarie}.

Since below we will present an advanced large-deviation algorithm which allows us to sample the distribution $Q(G)$ of cavity Green's function with a very high precision in the tails, beyond the scale ${\cal M}^{-1}$ set by the size ${\cal M}$ of the population, we need to specify explicitly the population dynamics approach~\cite{tikhcrit,PopDyn}: We store a population $\{G_l\}$ of ${\cal M}$ complex-valued elements $G_l=a_l+{\rm i} b_l$ ($l=1,\ldots, {\cal M}$), \ie, $a_l={\rm Re} (G_l)$ and $b_l={\rm Im} (G_l)$. For each iteration step, we pick $k$ randomly chosen elements $G_{l_j}=a_{l_j}+{\rm i} b_{l_j}$ from the population and draw a uniformly distributed random number $\epsilon$ for the local energy according to~(\ref{eq:peps}). This allows us to calculate a new element from~(\ref{eq:recursion}). 
Since below we will access the imaginary part of $G$ seperately, we use (\ref{eq:recursion}) in the following explicit form 
\begin{eqnarray}
a+{\rm i} b & = & \frac{\left(-\epsilon-E-\sum_{j=1}^k a_{l_j}\right)+{\rm i} 
\left(\sum_{j=1}^k b_{l_j}+\eta\right)}
{\left(-\epsilon-E-\sum_{j=1}^k a_{l_j}\right)^2+
\left(\sum_{j=1}^k b_{l_j}+\eta\right)^2} \nonumber \\ 
& \equiv & f_{E+\sum_{j=1}^k a_{l_j},\sum_{j=1}^k b_{l_j}+\eta}(\epsilon)\,,
\label{eq:anderson:local:field}
\end{eqnarray}
which implies the definition of $f_{A,B}(\epsilon)$ 
\begin{equation}
f_{A,B}(\epsilon)=\frac{(-\epsilon-A)+{\rm i}B}{(-\epsilon-A)^2+B^2}
\label{eq:transfer}
\end{equation}
for convenience. The iteration step is completed by replacing one randomly chosen element by the new one. This iteration is always performed until approximate convergence of the population, as established by monitoring mean, variance and few very small quantiles as well as the full shape of the distribution. Naturally, the resolution of the approximated distribution, represented by the population, is determined by the number of elements ${\cal M}$ in the population, as deeply investigated in~\cite{tikhcrit}.

Previous studies~\cite{abou,ourselves,tikhcrit} have shown that in the localized phase, $W > W_L \approx 18.2$ (in the ${\cal M} \to \infty$ limit), the iteration equations are unstable with respect to the imaginary regulator $\eta$: $Q(G)$ and $P({\cal G})$ are singular and the average DoS vanishes in the $\eta \to 0^+$ limit. Conversely, in the metallic phase the probability distributions converge to stable non-singular $\eta$-independent distribution functions, provided that $\eta$ is sufficiently small.

For the distribution $Q(b)$ of the imaginary part $b\equiv {\rm Im} G$ we aim at obtaining the distribution to a high precision, \ie, deep in the tails. For this purpose, we have implemented a large-deviation approach, which is explained next.
Standard large-deviation algorithms rely on sampling of biased distributions and unbiasing the obtained data in the end. Such approaches have been widely used, e.g., to study the large-deviation properties of random-graphs \cite{largest2011,diameter2018}, biological sequence alignments \cite{align2002}, protein folding \cite{dellago1998}, random walks \cite{fBm_MC2013,convex_hull2015}, models of transport
\cite{giardina2006,schreckenberg2019}, the Kardar-Parisi-Zhang equation \cite{kpz2018}, nonequilibrium work processes \cite{work_ising2014} and many more. We have tried such an approach based on a bias here, but were not able to see convergence of the used Markov chains deep enough in the tails. For this reason, we have developed a very different approach here.

To convey the main idea, we notice that for any given set of randomly selected elements $\{G_{l_j}\}$, the next (and only) step is to sample random energy values according to the uniform distribution to obtain the probability of the imaginary part $b$ conditioned to this set. This means, for the given set and given values of $E$ and $\eta$, corresponding to $A=E+\sum_{j=1}^k a_{l_j}$ and $B=\sum_{j=1}^k b_{l_j}+\eta$, we have, by using a standard property of the delta function $\delta(x)$ and by using that the probability density for the local energies is simply $1/W$,
\begin{eqnarray}Q_{A,B}(b) & = & \int_{-W/2}^{W/2}
\delta \left (b-\tilde f_{A,B}(\tilde \epsilon)\right) \frac 1 {W} d\tilde 
\epsilon
\label{eq:PAB}
\\
& = & \frac 1 {W} \int_{-W/2}^{W/2} \sum_{l: \tilde f_{A,B}(\tilde \epsilon_l)=b} 
\frac 1 {|\tilde f_{A,B}'(\tilde \epsilon_l)|} 
\delta(\tilde \epsilon - \tilde \epsilon_l) d\tilde \epsilon\, , \nonumber
\end{eqnarray}
where $\tilde f_{A,B}= {\rm Im} f_{A,B}$ and $\tilde \epsilon_l$ are those real-valued zeroes of $b-\tilde f_{A,B}(\tilde \epsilon)$ which are located in the interval $[-W/2,W/2]$, and $\tilde f_{A,B}'(\epsilon)$ is the derivative of $\tilde f_{A,B}$ with respect to $\epsilon$. The zeroes are simply to obtain, because we have to solve only a quadratic equation, leading to $\tilde \epsilon_l=A\pm \sqrt{B/b-B^2}$. 

Let us now assume that a arbitrary value $b$ is given (fixed), where we want to evaluate $Q(b)$. The requirement that we only have to consider real-valued roots leads immediately to $b\le 1/B$, \ie, $Q_{A,B}(b)=0$ for $b>1/B$.\footnote{This also follows directly from Eq.~(\ref{eq:anderson:local:field}) because the imaginary part  can be bounded from above by the value obtained for $(\epsilon-A)^2=0$.} This, on the other hand, means that to evaluate 
$Q(b)$, we could sample from the population such that only values are considered 
which follow this condition, \ie, where $B\le 1/b$, \ie, $\sum_{j=1}^k b_{l_j}+\eta \le 1/b$ holds. A simple way to achieve this restricted sampling is 
to sample values with $b_{l_j} \le 1/b-\eta >0$, since larger values will immediately lead to $Q(b)=0$. Still, because a sum $B=\sum_{j=1}^k b_{l_j}+\eta$ is calculated, sometimes the combined sample values will not meet the condition $B\le 1/b$, hence this gives rise no contribution to  $Q(b)$ as well. But this rejection happens much less frequently compare to sampling from the full distribution.\footnote{This could be improved even more by sampling the first element such that $b_{l_1}\le 1/b-\eta$, then sampling the second one such that  $b_{l_2}\le 1/b-\eta-b_{l_1}$ etc, but this would increase the efficiency only by an factor at most $k$ (here $k=2$), which we neglected, because the final sampling is anyway very fast, order of few seconds on a standard PC, as compared to the equilibration of the population, which takes more than one day.} Thus, we restricted the sampling of all $k$-tuples to the region $\le 1/b-\eta$ and included a bias $[\int_0^{1/b-\eta} \hat Q(\tilde b)d\tilde b]^k$ ($\hat Q$ is the approximation of the true probability as given by the finite population) to all values of $Q_{A,B}(b)$ as calculated from Eq.~(\ref{eq:PAB}). We technically  achieved the  restricted sampling by once sorting the population obtained in the standard population dynamics according to the value of the imaginary part $b_l$ and subsequently drawing uniformly inside the desired range. Note that if the ${l_{\max}}$'th element of the sorted population is the largest element which is inside the desired range, the bias is simply $(l_{\max}/{\cal M})^k$. For each value of $b$ we were interested in, we performed $N_{\rm est}$ times this step of estimating $Q(b)$   and averaged over these estimates. In Fig.~\ref{fig:algorithm} the algorithm is summarized.

\begin{figure}
\begin{tabbing} xx \= xx \= xx \= xx \= xx \= xx \= xx \= xxxxx \kill
 {\bf algorithm} sampling $Q(b)$\\
 {\bf begin}\\
 \> Initialize population of ${\cal M}$ members.\\
 \> Iterate population using Eq.~(\ref{eq:anderson:local:field})
     until convergence\\
 \> {\bf for} $b$ in desired range \\
 \> {\bf begin}\\
 \>\> $s=0$ \\
 \>\> {\bf for} t=1 to $N_{\rm est}$ \\
 \>\> {\bf begin}\\
 \>\>\> sample $k$ elements $\{g_{l_i}=a_{l_i}+{\rm i} b_{l_i}\}$ 
with $b_{l_i}\le 1/b-\eta$\\ 
 \>\>\> $A=E+\sum_{j=1}^k a_{l_j}$, $B=\sum_{j=1}^k b_{l_j}+\eta$ \\
 \>\>\> calculate $Q_{A,B}(b)$ according to Eq.~(\ref{eq:PAB})\\
 \>\>\> $s = s+ Q_{A,B}(b) \times [\int_{0}^{1/b-\eta} \hat Q(\tilde b) 
d\tilde b]^k$\\
 \>\> {\bf end}\\
 \>\> {\bf print} $b$, $s/N_{\rm est}$ \\
 \> {\bf end}\\
 {\bf end}\\
\end{tabbing}
\caption{Summary of the large-deviation sampling algorithm  for the
distribution of the imaginary part of the cavity Green's function (see text).
 \label{fig:algorithm}}
\end{figure}

\section{Results}
We have applied the large-deviation approach described above within computer simulations~\cite{practical_guide2015} to obtain the distribution of the cavity Green's function for the Anderson model for the Bethe lattice with degree $k+1=3$ with $E=0$ and $\eta=0$ for values of the disorder parameter $W\in [13,17.3]$. 
For the population dynamics approach, we use a population size ${\cal M}=10^7$ (for which, as  discussed in Ref.~\cite{tikhcrit} in great detail,  the transition point is expected to be shifted to slightly smaller value of the disorder compared to the ${\cal M} \to \infty$ limit). To speed up convergence, since the imaginary parts $b_l$ of the elements $G_l$ are typically small with increasing value of $G$, we initalized the elements with random values uniformly distributed for the real parts as $a_l \sim U(-1,1)$ and for the imaginary parts as $b_l \sim 10^{-\delta}U(0.5,1.5)$. We used $\delta=0$ (no special scaling) for $W\le 16$ and $\delta={9}$ for $16<W\le 17.4$. For all values of $W$, we observed convergence when iterating the population $10^4$  times (\ie, $10^4 \times {\cal M}$ times Eq.~(\ref{eq:anderson:local:field}) is evaluated). For the final estimate of $Q(b)$ we used $N_{\rm est}=10^4$ and considered logarithmically spaced values of $b\ge 1$.

\begin{figure}[ht]
\includegraphics[width=0.48\textwidth]{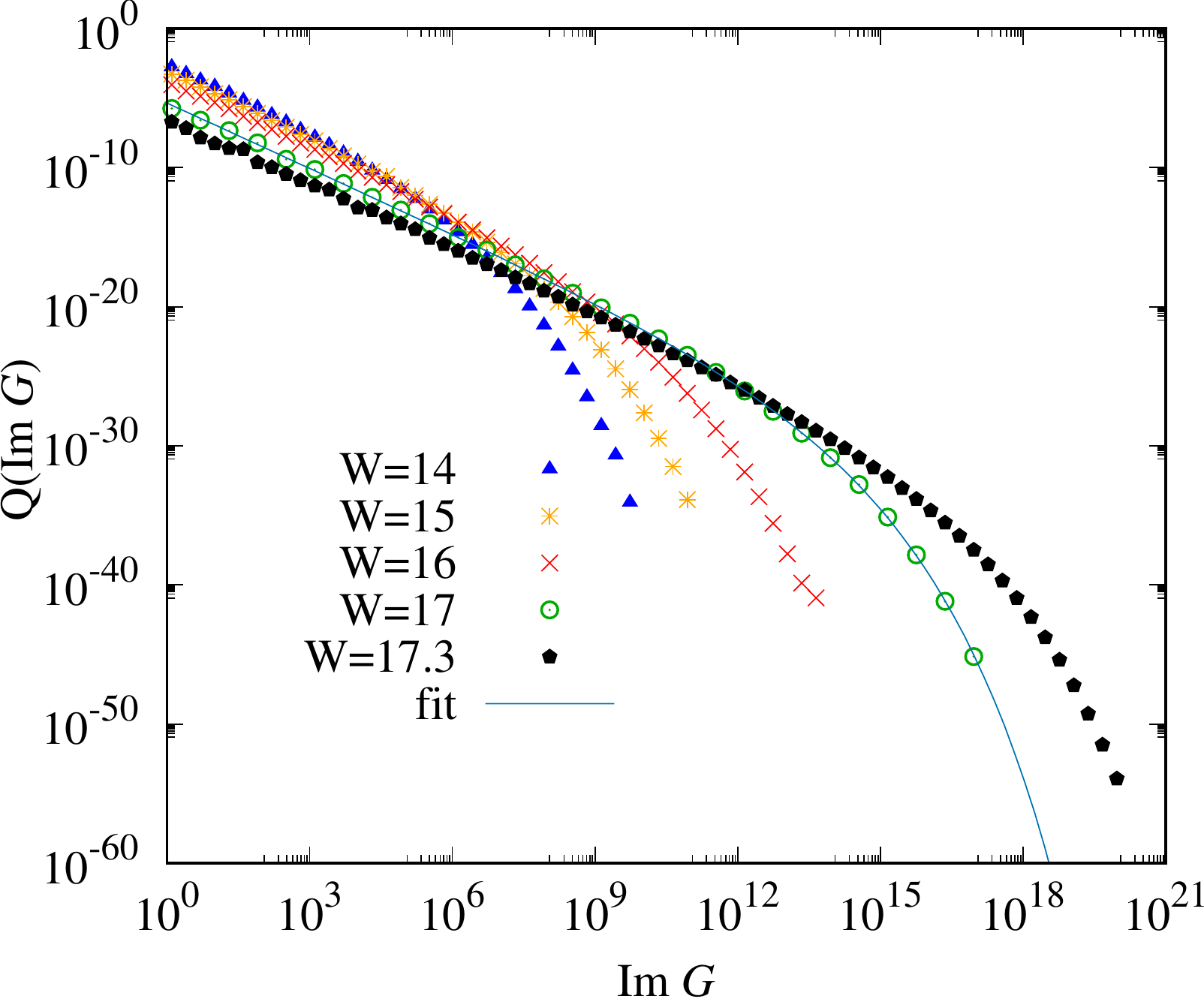}
\caption{(color online) Distribution $Q({\rm Im} G)$ of the ${\rm Im} G$ of cavity Green's function for some values $W\in[13,17.3]$. The line shows the result of a fit according to Eq.~(\ref{eq:shape:distr}) to determine the correlation volume, see text. 
\label{fig:distr_g_neu}}
\end{figure}

The resulting distributions $Q(b)$ for the imaginary part $b= {\rm Im} G$ is shown in Fig.~\ref{fig:distr_g_neu}. Note that using the large-deviation approach, probability densities as small as $10^{-50}$ can be accesses with a very high precision, well below any probability reached by a standard population dynamics approach. To extract the correlation volume, we assume that the distribution follows the heuristic shape
\begin{equation}
f(b) = f_0 b^{-\lambda} \exp\left( -(b/N_c)^{\alpha}\right)
\label{eq:shape:distr}
\end{equation}
where the behavior for small values of ${\rm Im} G$ is governed by a power decay with exponent $\lambda$ and the tail behavior by a stretched exponent with exponent $\alpha$ and scale $N_c$. 

Note that we also tried the fitting from given in Eq.~(57) of Ref.~\cite{mirlin1994}, corresponding of the analytic prediction of the supersymmetric treatment for the asymptotic behavior of the tails of the probability distribution close to $W_L$. Such function fits equally well the data of Fig.~\ref{fig:distr_g_neu}. However, it contains trade-off parameters for the correlation volume, 
\ie, it is possible to obtain good fits to the tail of the distributions over broader ranges of $N_c$ for suitably chosen combinations of the values of the other parameters. Therefore, in order to obtain a more informative estimation of the correlation volume, we finally only considered Eq.~\eqref{eq:shape:distr}. 

\begin{figure}[ht]
\includegraphics[width=0.48\textwidth]{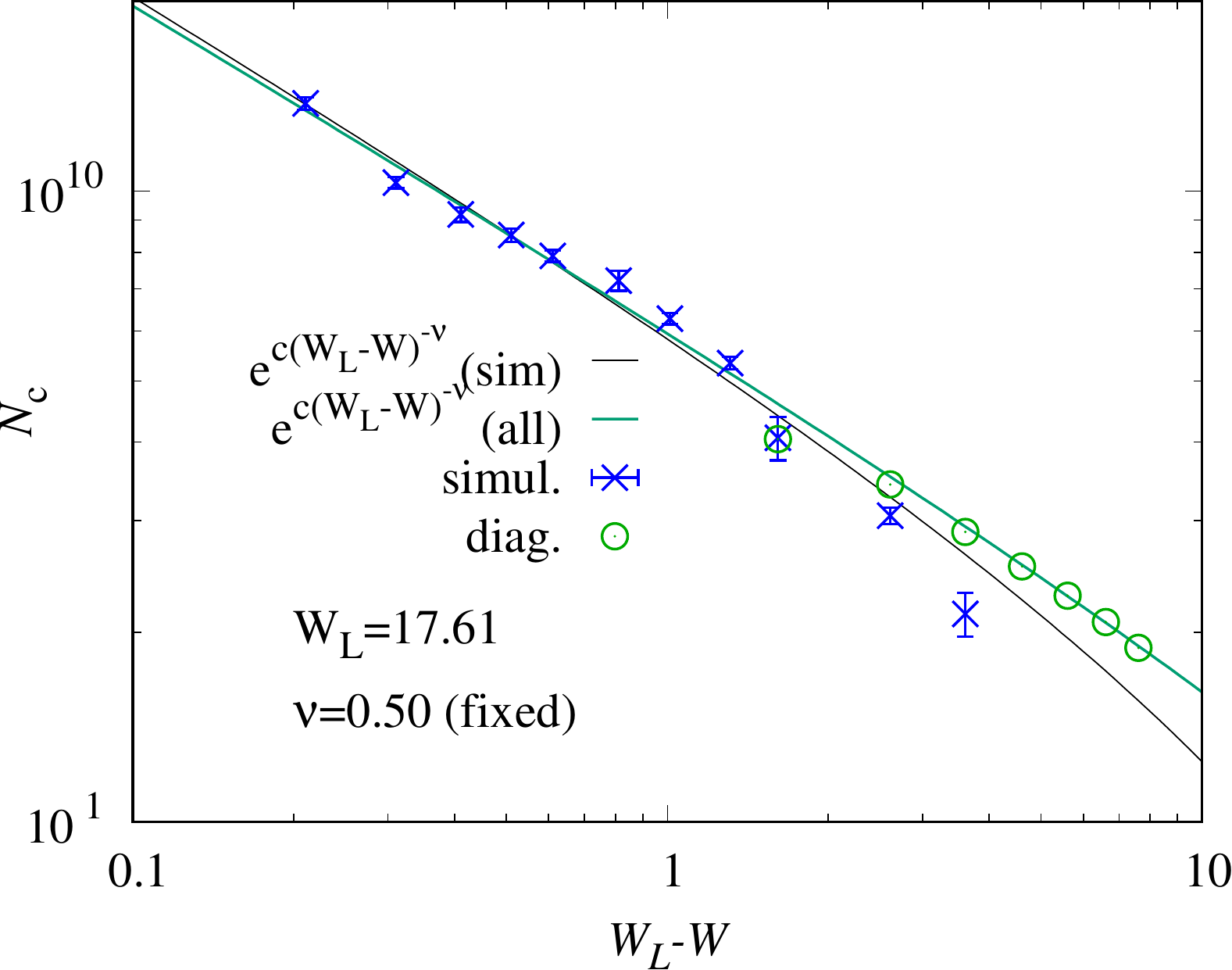}
\caption{(color online) Log of the correlation volume $N_c$ as a function of the distance of the disorder parameter $W$ from the critical point $W_L$, as obtained from the cut-off of the tails of $Q({\rm Im} G)$ using the large-deviation approach (blue crosses), and from the non-monotonic behavior of $q^{\rm typ}$ using EDs (turquoise circles, corresponding to the turquoise circles of Fig.~\ref{nc}). The lines show the result of fits which model the divergence of the scale at $W_L$ according to Eq.~(\ref{eq:diverging:scale}). The upper line is when fitting the large-deviation data only, while the lower line is for all data combined, which results in $W_L=17.61(3)$. Just here, since the data is plotted as function of $W_L-W$, a single value of $W_L$ is needed to see a power-law behavior for $W\to W_L$. Thus, for fitting the large-deviation data, the fixed value of the same $W_L=17.61$ was used. Note that  when $W_L$ is allowed to adjust here, a similar value $W_L=17.77(8)$ results, which is the value mentioned in the text. 
\label{fig:lengthscale_g}}
\end{figure}

By fitting the (log of the) distributions using the heuristic function~\eqref{eq:shape:distr} for the different values of $W$, we obtained the cut-off scale as a function of disorder strength $W$. Note that for $\lambda$ we obtained values near $1.5$, compatible with the prediction of~\cite{mirlin1994,tikhonov2019}. We thus fixed $\lambda=1.5$ for all values of $W$, resulting in less noisy data for $N_c$ for the final fits. For the exponent $\alpha$, we obtained values in the range $\alpha \in [0.163(2),0.208(2)]$
with a decreasing trend for growing values of $\lambda$. The results for $N_c$ are shown in Fig.~\ref{fig:lengthscale_g}.  We also show on the same plot the estimation of the correlation volume extracted from the non-monotonic behavior of $q^{\rm typ}$ obtained via EDs (circles of Fig.~\ref{nc}, see also Fig.~\ref{minimo2}). This comparison if very insightful for two reasons. (i) The largest correlation volume obtained using the large-deviation approach for $W=17.3$ is about $5.8 \times 10^{13}$ which is almost $2^{46}$. Thus, to observe such correlation volumes directly using EDs, one would have to treat RRGs of at least this size, which is, comparing to the results shown in section \ref{sec:numerics}, clearly infeasible with current methods. (ii) The estimations of $N_c$ obtained from the non-monotonic behavior of the spectral statistics and from the cut-off of the tails of the probability distribution of the LDoS 
can have a different prefactor $A$ appearing in Eq.~(\ref{eq:diverging:scale}), and asymptotically coincide only close enough to the Anderson transition. Far for the transition the two estimations can lead to quite different results. 
 Yet, Fig.~\ref{fig:lengthscale_g} shows that the two estimations of $N_c$ are in surprisingly good agreement even far below $W_L$.

We have fitted the resulting scale values to the function
\begin{equation}
N_c(W) = A \, e^{c/(W_L-W)^\nu}
\label{eq:diverging:scale}
\end{equation}
(actually by fitting $\log N_c(W)=\log A + c(W-W_L)^\nu$ to $\log$ of the measured scale). When setting $\nu=0.5$, we obtained estimates $W_L=17.77(8)$ and $c= 21(3)$ (just statistical error bars). We also tried to fit with the same function 
the combination of the large deviation data and the data from the EDs extracted from the non-monotonic behavior of $q^{\rm typ}$, and we got $W_L=17.61(3)$ and $c=15(1)$,
see also Fig.~\ref{fig:lengthscale_g}. 

\section{Conclusions and perspectives} \label{sec:conclusions} 

In this paper we have introduced a new large-deviation approach to investigate the critical behavior of the Anderson model on the RRG. This approach allows us to study the distribution of the imaginary part of the cavity Green's function down to very small probability tails which are completely out of reach for standard numerical techniques.

In fact, as shown in Sec.~\ref{sec:numerics} and previously discussed in Refs.~\cite{mirlin,levy,biroli2018}, EDs clearly indicate the existence of a characteristic crossover scale $N_c(W)$ governing the finite-size effects of several observables and probes associated to the statistics of the gaps and of the eigenfunctions' amplitudes: For small sizes $N \ll N_c$ these observables seem first to flow towards towards the critical value upon increasing $N$  (which on the RRG correspond to the ones of the localized phase~\cite{largeD,fyod,efetov,Zirn}), and then for $N \gg N_c$ eventually approach the values corresponding to a standard delocalized, fully ergodic, phase. Although the ED estimation of $N_c(W)$ is compatible with an exponential divergence of the correlation volume upon approaching the Anderson transition, the numerical data are limited to relatively small sizes, $N \le 2^{15}$, and thus can only access a disorder range too far from the transition to allow one for an accurate determination of its critical behavior. 

On the contrary, the large-deviation extension of the population dynamics approach allowed us to obtain accurately the distribution of the imaginary part of the cavity Green's function to very small probability densities as $10^{-50}$ (in order to obtain them by ED one would need a system size at least as large as $N=2^{46}$ sites). The main idea, is to first perform a standard population dynamics till convergence. In a second step, a biased sampling of  the such obtained histogram is made. This works out, because for given values of ${\rm Im} G$, only a restricted range of the histograms contributes, and the magnitude of this range determines the bias used. 

To extract the correlation volume $N_c$, we have fitted the distributions by using  a stretched exponential, which describes very well the data. Our result provides the strongest and more direct numerical evidence to date of a divergence of the logarithm of the correlation volume with a power of $\nu=0.5$~\cite{SUSY,fyod,mirlin1994,tikhonov2019,tikhcrit,lemarie,Zirn}. The corresponding transition value that we find is $W_L\approx 17.77$. 

These results provide another transparent and coherent argument supporting the idea that the Anderson model on the RRG becomes fully ergodic in the whole delocalized phase, in agreement with the recent results of~\cite{mirlin,levy,lemarie,biroli2018} and with the predictions of~\cite{tikhonov2019,SUSY,fyod,mirlin1994,Zirn} based on supersymmetric field theory. Nonetheless, ergodicity establishes on a system size which becomes exponentially large as the localization transition is approached, and exceeds the system sizes accessible via ED well before the localization transition, resulting in a very wide crossover region in which the system looks as if it were in a mixed (delocalized but non-ergodic) phase for all practical purposes, \ie, on finite but large 
length and time scales (volumes smaller than $N_c(W)$ and times smaller than $\hbar/N_c^{-1} (W)$).

In the light of the analogy between Anderson localization on Bethe lattices and  Many-Body Localization~\cite{A97,BAA,jacquod,wolynes,scardicchioMB,roylogan,mirlinreview,dinamica}, the results presented above might help us understand the highly non-trivial properties of the delocalized phase of many-body interacting disordered systems exhibiting MBL.

On the methodological side, our approach might present a new type of large-deviation approach. It could be helpful also for other models, where similar self-consistent equations, like for field distributions, are obtained. 
The key property is Eq. (\ref{eq:PAB}), which gives the contribution to an arbitrary location $b$ of the desired distribution $Q$ as function of any given sample of the previously obtained population and as a function of the underlying disorder distribution.
In case the sampling of the population can be effectively restricted to the relevant values, depending on $b$, and if the zeros of the delta function can be obtained efficiently, our proposed approach should be useful. 

\begin{acknowledgments}
 This research was partially supported by a grant from the Simons Foundation ( \# 454935 Giulio Biroli). We thank the Centre National de la Recherche Scientifique for supporting AKH during a guest professorship at Sorbonne Universit\'e. 
   The simulations were performed at the HPC Cluster CARL, located at the University of Oldenburg (Germany) and funded by the Deutsche Frschungsmemeinschaft (DFG) through its
  Major Research Instrumentation Program (INST 184/157-1 FUGG) and the Ministry of  Science and Culture (MWK) of the Lower Saxony State.
\end{acknowledgments}


\end{document}